# Improved topological conformity enhances heat conduction across metal contacts on transferred graphene

Bin Huang[1], Yee Kan Koh[1]

[1]Department of Mechanical Engineering, and Centre for Advanced 2D Materials, National University of Singapore, Singapore

## ABSTRACT

Thermal conductance of metal contacts on transferred graphene (trG) could be significantly reduced from the intrinsic value of similar contacts on as-grown graphene (grG), due to additional resistance by increased roughness, residues, oxides and voids. In this paper, we compare the thermal conductance ($G$) of Al/trG/Cu interfaces with that of Al/grG/Cu interfaces to understand heat transfer across metal contacts on transferred graphene. Our samples are polycrystalline graphene grown on Cu foils by chemical vapor deposition (CVD) and CVD-grown graphene transferred to evaporated Cu thin films, both coated with a thin layer of Al. We find that for the Al/grG/Cu interfaces of as-grown CVD graphene, $G$=31 MW $m^{-2}$ $K^{-1}$ at room temperature, two orders of magnitude lower than that of Al/Cu interfaces, due to weak coupling of electrons in the metals and graphene. For most as-transferred graphene on Cu films with root-mean-square (rms) roughness of 2-6 nm that we measured, $G\approx$20 MW $m^{-2}$ $K^{-1}$, $\approx$35% lower than that of as-grown CVD graphene. We carefully rule out the contributions of residues, native oxides and interfaces roughness, and attribute the difference in the thermal conductance of as-grown and as-transferred CVD graphene to different degrees of conformity of graphene to the Cu substrates. We find that a contact area of $\approx$50% only reduces the thermal conductance by $\approx$35%, suggesting that a small but measurable amount of heat transfer occurs across "voids" at graphene interfaces. We successfully improve the conformity of the as-transferred graphene to the substrates by annealing the samples at 300°C, and thus enhance the thermal conductance of the transferred graphene to the intrinsic value. From the temperature dependence measurements of $G$ of Al/trG/Cu and Al/grG/Cu interfaces, we also confirm that lattice vibrations (i.e., phonons) are the dominant heat carries across the metal/graphene/metal interfaces despite a substantial carrier concentration of ≈3×10$^{12}$ cm$^{-2}$ induced in the graphene.

**TEXT**

Due to high in-plane thermal conductivity of graphene,[1-3] heat transfer through graphene interfaces plays a pivotal role in thermal management of graphene devices. For example in electronic and optoelectronic devices, hot spots are usually formed in graphene due to self-heating as the devices are driven by high electric fields while operating.[4-6] Depending on lateral sizes of graphene, heat dissipation from these hot spots in graphene devices is dominantly governed by the thermal resistance of either graphene/substrate interfaces (when graphene is larger than ≈500 nm) or graphene/metal contacts (when graphene is smaller than ≈500 nm).[7] In a similar manner, the proposed applications of graphene as efficient heat spreaders,[8,9] flexible thermal heaters[10] and fillers to enhance the performance of thermal interface materials[11] also rely on low thermal resistance between graphene and the underlying substrate.

Despite the importance, knowledge of heat transfer across graphene interfaces is still relatively limited. Prior measurements on graphene interfaces mostly focus on two readily achievable interfaces (i.e., metal/graphene/SiO$_2$[7, 12-14] or graphene/SiO$_2$[15-17]) of either mechanically exfoliated graphene or transferred CVD graphene, while a wider range of graphene interfaces has been studied theoretically.[18-22] Although these early studies are crucial milestones in understanding heat transport across graphene interfaces, critical questions still remain unresolved. For example, prior measured $G$ of graphene interfaces is relatively low (20-100 MW m$^{-2}$ K$^{-1}$) compared to the intrinsic values of *G* of epitaxial solid/solid interfaces (≈600 MW m$^{-2}$ K$^{-1}$).[23] This low value of thermal conductance was attributed to a high mismatch between energy of phonon modes in graphene and that in substrates or metals.[7] However the low value of *G* could also be partially contributed by reduction of *G* from the intrinsic values due to imperfect quality of the interfaces after graphene transfer. Despite impressive progresses in clean transfer[24-26] of graphene, it is well documented that the transferred graphene is imperfect with wrinkles,[27] ripples and polymer residues.[25] These imperfections could significantly retard heat flow across interfaces of transferred graphene. Moreover, while graphene generally conforms to the substrates,[28, 29] the degree of conformity of graphene was not characterized in previous studies of heat transport across graphene interfaces and hence the existence of voids cannot be discounted. Theoretical study suggests that heat transfer only occurs through areas of real physical contact,[21] but this is yet to be verified experimentally since a substantial amount of heat could flow through the voids at the interfaces if they are filled by water and hydrocarbons.[7] Lastly, graphene interfaces present a unique material system to study the effects of interfacial roughness to heat conduction across interfaces, since graphene interfaces are chemically abrupt without intermixing of atoms. Prior measurements on Al/Si interfaces[30, 31] suggest that *G* is reduced by interfacial roughness and native oxide layers on underlying substrates. On the contrary, a recent molecular dynamics study suggests that *G* of graphene/Cu interfaces is significantly enhanced by increased interfacial roughness, due to enhanced coupling between graphene and copper.[20] It is thus still unclear whether interfacial roughness enhances or reduces heat flow across interfaces.

In this paper, we provide new insights into these critical questions by measuring the thermal conductance of interfaces of as-grown and transferred CVD graphene on Cu films/foils. We find that the intrinsic value of thermal conductance of Al/grG/Cu interface is relatively low; $G$=31 MW m$^{-2}$ K$^{-1}$ for as-grown CVD graphene (grG) on copper foil coated with Al. For most transferred graphene (trG) samples, we derive $G\approx$20 MW m$^{-2}$ K$^{-1}$, ≈35% lower than $G$ of interfaces of as-grown graphene. Surprisingly, we find that $G$ is independent of interfacial roughness for root-mean-square (rms) roughness up to 6 nm. From the atomic force microscopy (AFM) topographic images of these as-transferred graphene, we estimate a percentage of contact areas of only ≈50%, suggesting that a small but measurable amount of heat is transferred through the voids at the interfaces. By annealing as-transferred graphene samples at 300°C, in either forming gas for 3 hours or vacuum for 30 mins, we improve the percentage of contact areas of the annealed graphene to ≈100% and thus achieve the intrinsic value of $G\approx$31 MW m$^{-2}$ K$^{-1}$ for the annealed samples. Our results provide comprehensive understanding of thermal transport across interfaces of as-grown and transferred graphene.

Our samples are three sets of Al/graphene/Cu samples. Set A consists of as-grown graphene (grG) on copper foils purchased from Graphene Supermarket. The graphene was grown by chemical vapor deposition (CVD), with the unintended graphene on the opposite side of the copper foils being etched away by reactive ion etching. Set B consists of as-transferred graphene (trG) samples. The CVD-grown graphene from Graphene Supermarket was transferred on a series of copper films on 100 nm thick SiO$_2$ on Si. The Cu films are approximately 0.2, 0.4 and 1 μm thick, deposited by thermal evaporation with a deposition rate of 2-20 Å/s and a base pressure of <10$^{-8}$ Torr. By changing the deposition rate and the thickness of the copper films, we achieved copper films with rms roughness of $\eta$=2.2 nm, 3.5 nm and 6 nm, measured by atomic force microscopy (AFM). Set C is a series of annealed graphene samples, with the transferred graphene on the evaporated copper films annealed at 300°C using a tube furnace, either in vacuum (10$^{-4}$ Torr) for 30 mins or in forming gas (95% N$_2$, 5% H$_2$) for 3 hours. The annealing improves the conformity of graphene to the substrates, and thus enhances the thermal conductance of Al/graphene/Cu interfaces, as explained below.

We deposited a ≈100 nm thick Al film on all samples by thermal evaporation. The Al films act as the transducer for our thermal measurements.

We follow procedures listed in Ref. 25 to perform clean graphene transfer. We use poly(bisphenol A carbonate) (PC) with a molecular weight of 45000 from Sigma Aldrich to prepare a 1.5 wt. % PC in chloroform solution, to be spin-coated onto graphene as the support layer for the transfer. We choose PC instead of the more common poly(methyl methacrylate) PMMA because it is easier to dissolve PC residues on the transferred graphene due to the smaller molecule size and weaker interfacial adsorption.[25] With a spin-coated PC thin film of ≈100 nm, we float the graphene samples on a 7 wt. % ammonium persulfate (APS) solution to etch away the underside copper. We clean our graphene using a modified RCA process[32] after the etching, transfer the graphene to the substrates and bake the samples to remove the excess water. We then soak the samples in chloroform for 24 hours to remove the PC layer. Lastly, we rinse the samples in isopropanol alcohol (IPA) and blow dry the samples using dry nitrogen.

We employed Raman spectroscopy, atomic force microscopy (AFM) and x-ray photoelectron spectroscopy (XPS) to ensure that the graphene is clean and undamaged after the transfer process. We measured the Raman spectra of both epitaxial and transferred graphene using a home built Raman system with a 532 nm continuous wave (cw) laser. We find that the graphene sheets are monolayer and undamaged with no significant D peaks, see Fig. 1b. The G peak, however, is red-shifted to 1589 $cm^{-1}$, due to Landau damping and stiffening of phonons by charge carriers in graphene.[33] From the magnitude of both the red shift of G-peak and blue shift of 2D-peak, we estimate that a carrier concentration of ≈$3\times10^{12}$ $cm^{-2}$ is induced in graphene as a result of electron transfer from Cu.[33] The annealing in vacuum and forming gas does not change the position of G peak and thus does not alter the induced carrier concentration. We also do not observe a significant D peak in the Raman spectrum of a graphene/SiO$_2$ sample coated with a ≈7 nm thick Al film evaporated with the same deposition rate, see Fig. S1d in the online supporting materials; thus the thermal evaporation does not damage the graphene. We obtained the surface topography of the samples by AFM in a tapping mode. The AFM images indicate that the transferred graphene is clean and with minimum residues, see Fig. S1b of the online supporting materials. We also quantified the amount of residues and contaminants on the transferred graphene by XPS. We find that the ratio of the integrated intensity of the C=O peak (which we attribute to contaminants) and that of the peak of sp$^2$ hybridized carbons is <0.07, see Fig. 1c, comparable to the ratio we obtained for as-grown graphene on Cu foils. This confirms that the amount of PC residues is negligible.

We measured the thermal conductance $G$ of Al/graphene/Cu interfaces by time-domain thermoreflectance (TDTR).[23] A schematic diagram of our setup is presented in Ref. 46. In TDTR measurements, a train of ultrashort laser pulses from a Ti:sapphire laser is split into a pump and a probe beams by a polarizing beam splitter. The pump beam is modulated by an electro-optic modulator (EOM) at a radio frequency (rf) of 10 MHz. The modulated pump beam is absorbed by the Al films of the samples and thus creates a periodic heating at the surface of the samples. During the experiments, the relative time between pump and probe pulses, usually called the delay time $t_d$, is adjusted by changing the optical path of the pump beam using a mechanical stage and a retroreflector. The periodic temperature response at the sample surface is then monitored as a function of the delay time $t_d$, via the change of reflectance of the probe beam with temperature (i.e., thermoreflectance), using a photodiode and a rf lock-in amplifier. For current studies, we used laser 1/e$^2$ radii of 6 μm and a total laser power of 60-90 mW, to limit the steady-state temperature rise to <10 K.

Since heat dissipation from sample surface depends on thermal properties of the samples, we accurately derived the thermal conductance $G$ of the Al/graphene/Cu interfaces from our TDTR measurements. In the analysis, we compare the ratios of in-phase and out-of-phase signals of the rf lock-in amplifier to calculations of a thermal model[34] that analytically solves the heat flow in layered structures, see Fig. 1d. By analyzing the ratios instead of the absolute values of the measurements, we eliminate the need to accurately measure the laser power absorbed in the experiments. During the fitting, we determine most parameters in the thermal model from either the literature or independent measurements. We measured the thicknesses of the Al, Cu, and $SiO_2$ layers of our samples by picosecond acoustics, with an uncertainty of <3%. We estimated the thermal conductivity of the Al and Cu films from the electric resistivity of the evaporated films, measured by a four-point probe, using the Wiedemann-Franz law. We then adjusted the thermal conductance $G$ as the only free fitting parameter in the thermal model until calculations and measurements agree, see Fig. 1d. The uncertainties of the derived thermal conductance of the Al/graphene/Cu interfaces, for the as-grown and as-transferred graphene, are estimated to be ≈6 % and ≈10 %, respectively.

We find that $G$=31 MW m$^{-2}$ K$^{-1}$ at room temperature for the Al/grG/Cu interfaces of as-grown CVD graphene (grG). This value of thermal conductance is comparable to the thermal conductance of exfoliated graphene (exG) of Au/exG/$SiO_2$ interfaces,[7] but is two orders of magnitude smaller than the thermal conductance of Al/Cu interface[35] without the CVD graphene of ≈4 GW m$^{-2}$ K$^{-1}$. This low value of thermal conductance suggests that heat transfer by electrons is negligible across the metal/graphene/metal interface due to e.g., weak coupling between d-orbitals of Cu and Al and the π-orbitals of graphene.[36]

For the transferred graphene (trG), however, we obtain two distinct values for the thermal conductance of the Al/trG/Cu interfaces, see the dashed lines in Fig. 2. For the majority of the as-transferred samples (Set B), we find that $G \approx 20$ MW m$^{-2}$ K$^{-1}$, $\approx 35\%$ lower than the intrinsic value of $G$ of the Al/grG/Cu interfaces. However, for some as-transferred samples, we obtain the intrinsic value of $G \approx 31$ MW m$^{-2}$ K$^{-1}$. Interestingly, for all the transferred graphene samples annealed at 300°C (Set C), we find that $G \approx 31$ MW m$^{-2}$ K$^{-1}$, irrespectively of the duration of the annealing (30 mins and 3 hours) and whether the annealing is performed in vacuum or in forming gas. We stress that the differences we observed are not due to varying amounts of PC residues after the transfer. We observe no significant residues from the AFM phase images that we took after the transfer, see Fig. S3 in the online supporting materials for the examples of the AFM phase images of our samples. Moreover, XPS spectrum of the transferred graphene also indicates an insignificant amount of PC residues, see Fig. 1(c).

To understand the origins of the observed differences in $G$ of the interfaces of transferred graphene, we evaluate the impacts of interfacial roughness and the role of copper oxide on $G$ of Al/trG/Cu, see Figs. 2(a) and 2(b). We modified the root-mean-square (rms) roughness ($2 \leq \eta \leq 6$ nm) of the deposited copper films by varying the evaporation rate (2-20 Å/s) and film thickness (0.2-1 μm) during thermal evaporation of the copper films. We derived the rms roughness of our samples before and after graphene transfer from the AFM images and find that some of the samples become less rough after the transfer. We then plot the thermal conductance $G$ of the epitaxial and transferred graphene as a function of rms roughness $\eta$ of the graphene samples, see figure 2(a). We observe no significant dependence of $G$ on $\eta$. This finding is in contrast with prior measurements of Al/Si interfaces[30] and a recent molecular dynamics prediction[20] that the thermal conductance depends strongly on the interfacial roughness. One possible explanation to the lack of dependence on $\eta$ of graphene interfaces is that most heat-carrying phonons are already scattered at the interfaces when $\eta=2$ nm,[37] considering the high acoustic impedance mismatch between graphene and the metals, and that most heat-carrying phonons have a wavelength of <1 nm at room temperature.

To evaluate the roles of native copper oxide in heat transfer across the graphene interfaces, we measured the thickness ($h_{CuOx}$) of the native copper oxide on the surface of the copper films/foils of our as-grown, as-transferred and annealed graphene samples by XPS.[38] We find that $h_{CuOx}$=0.5 nm and 2.5 nm for the as-grown and as-transferred graphene samples, respectively, while annealing the transferred graphene samples in forming gas for 3 hours reduces $h_{CuOx}$ to ≈2 nm. We plot $G$ of the Al/G/Cu interfaces in Fig. 2(b) and find that $G$ is independent of $h_{CuOx}$. This result is consistent with the fact that the thermal resistance of a 2.5 nm thick $CuO_x$ film (≈$2.5 \times 10^{-9}$ m$^2$ K W$^{-1}$) is negligible compared to the measured thermal resistance ($1/G$≈$3.3 \times 10^{-8}$ m$^2$ K W$^{-1}$) of the graphene interfaces. Also, the lack of dependence on $h_{CuOx}$ supports our previous conclusion that contribution of electrons to heat transport across Al/grG/Cu interfaces is negligible even for as-grown graphene. If electronic heat transfer were significant, a 2.5 nm thick copper oxide should be sufficient to substantially reduce the contribution by electrons, and thus result in a measurable reduction in $G$.

We hypothesize that the observed differences in $G$ are due to different degrees of conformity of our as-transferred and annealed graphene samples. We examine the topography of the AFM images and notice that the topography of the transferred graphene with low thermal conductance has elongated ridges-like morphology, which is different from that of bare copper film with more uniform circular-like features, see Fig. 3. To quantify the conformity of graphene to the Cu films, we derived depth histograms[29] from the AFM topographic images, and compare the depth histograms of our samples before and after the graphene transfer. In our depth histograms, instead of plotting the more common percentage of counts (ϕ), we define a spatial frequency ξ=∂ϕ/∂h with a unit of reciprocal nanometer (nm$^-$

$^1$), where $h$ is the height measured by AFM. Using $\xi$ instead of $\phi$, our depth histogram does not depend on the arbitrary size of the intervals of height and we always have $\int_{-\infty}^{\infty}\xi dh=1$. The depth histograms of our as-transferred and annealed graphene samples indicate that not all graphene follows the topography of, and thus conforms to, the underlying Cu film, see Fig. 3 for a comparison of depth histograms of an as-transferred and an annealed graphene samples. We then fit the depth histograms with a Gaussian function and derived the standard deviation of the height distribution from the depth histograms of the transferred graphene ($\sigma_g$) and the corresponding Cu thin film ($\sigma_{Cu}$). We observe that $\sigma_g<\sigma_{Cu}$, indicating that graphene is partially suspended atomic "peaks" of Cu thin films. We thus define $\Delta\sigma=\sigma_{Cu}-\sigma_g$ to quantify the degrees of conformity;[39] $\Delta\sigma=0$ if graphene fully conforms to the Cu thin film.

We plot $G$ of the Al/trG/Cu interfaces as a function of the derived $\Delta\sigma$ in Fig. 2(c). We find that for all conformal graphene (i.e., $\Delta\sigma\approx0$), $G\approx30$ MW m$^{-2}$ K$^{-1}$, while for all non-conformal graphene, $G\approx20$ MW m$^{-2}$ K$^{-1}$, see Fig. 2(c). Analysis of the depth histograms of our annealed graphene samples suggests that annealing graphene either in vacuum for 30 mins or in forming gas for 3 hours is sufficient to increase the conformity of the as-transferred graphene and thus enhance $G$ of the annealed graphene to the intrinsic value.

To assess whether heat flows through voids at graphene interfaces, we estimate the percentage of contact areas at the interfaces of transferred graphene from the depth histograms. To do so, we plot the heights of the individual pixels in the AFM images of the Cu substrate and the graphene, in a descending manner, as a continuous smooth function of the accumulated counts, see Fig. S4 in the online supporting materials. We assume that graphene conforms to the asperities on Cu and adjust the absolute heights of the graphene and Cu to match at top 5%. We consider graphene to be in contact with Cu if the height difference is <0.5 nm, roughly the height of a monolayer on a substrate.[40] We find that the percentages of contact areas are ≈100 % for conformal samples and ≈50 % for non-conformal samples, respectively. By plotting $G$ of Al/trG/Cu interfaces as a function of percentage of contact areas, we observe that $G$ does not scale linearly with the percentage of contact areas, see Fig. 2(d). Our measurements thus suggest that a small but measurable amount of heat (≈10 %) of the heat is carried across the voids at the interfaces of the non-conformal samples, see Fig. 2(d).

Our claim of different degrees of conformity for the interfaces of as-transferred and annealed graphene is further supported by the acoustic echoes[41] due to the graphene interfaces, measured in-situ during the TDTR measurements by picosecond acoustics, see Fig. S5 of the online supporting materials. We find that there is a significant difference in the shapes of the second echoes between the non-conformal as-transferred graphene and conformal annealed graphene, see Fig. S5a. Furthermore, we plot the amplitude of the acoustic echoes, normalized by the laser power used in the experiments, as a function of the percentage of contact areas, see Fig. S5b. We find that the acoustic echoes are weaker when graphene conforms.

Finally, we plot $G$ of Al/graphene/Cu interfaces of an as-grown graphene, an as-transferred graphene and an annealed graphene over a temperature range of $80 \leq T \leq 600$ K in Fig. 4. The lack of dependence on temperature for $T > 200$ K, typically found in other graphene interfaces, see Fig. 4, supports our previous conclusion that heat is mainly carried by lattice vibrations (i.e., phonons) and not by charge carriers, despite a relatively high carrier concentration of $\approx 3 \times 10^{12}$ cm$^{-2}$. In Fig. 4, we also plot the electronic thermal conductance $G_e$ estimated from the electrical specific contact resistivity of the Al/graphene/Cu interface $\rho_c$; $G_e = LT/\rho_c$,[31] where $L = 2.45 \times 10^{-8}$ Ω W K$^{-2}$ is the Lorenz number. Since $\rho_c$ of Al/graphene/Cu interface is not available in the literature, we assume that $\rho_c$ is independent of temperature and approximate $\rho_c=10^{-7}$ Ω cm$^{-2}$ from the specific contact resistivity of a series of metal/graphene contacts.[42] We note that the actual value of $\rho_c$ could be higher due to enhanced bonding in the Al/graphene/Cu sandwich structure.[43] Using this value of $\rho_c$, we approximate $G_e=0.7$ MW m$^{-2}$ K$^{-1}$ at 300 K, two orders of magnitude smaller than the measured $G$.

In summary, we report the intrinsic value of the thermal conductance of Al/grG/Cu interface of as-grown CVD graphene; $G\approx30$ MW m$^{-2}$ K$^{-1}$. We find that the thermal conductance of as-transferred graphene could be significantly reduced from this intrinsic value if graphene does not conform fully to the Cu substrate. We quantified the percentage of contact areas for our non-conformal graphene from the AFM topographic images, and find that the thermal conductance is reduced by ≈35% when the contact areas are about 50%. Our results thus suggest that a measurable amount of heat is carried through the voids at graphene interfaces, probably because the voids are not empty but filled with water vapor and hydrocarbon. We successfully recover the intrinsic value of thermal conductance by annealing the non-conformal graphene at 300°C. We also performed temperature dependence measurements on our as-grown, as-transferred and annealed graphene samples. We confirm that heat is carried by phonons in all these samples. Thus, charge carriers do not contribute to heat conduction across the metal/graphene/metal interfaces despite a substantial concentration of charge carriers induced in graphene by the metal contacts.

## Acknowledgements

This work is supported by NUS Young Investigator Award 2011. Sample characterization was carried out in part in the Centre for Advanced 2D Materials.

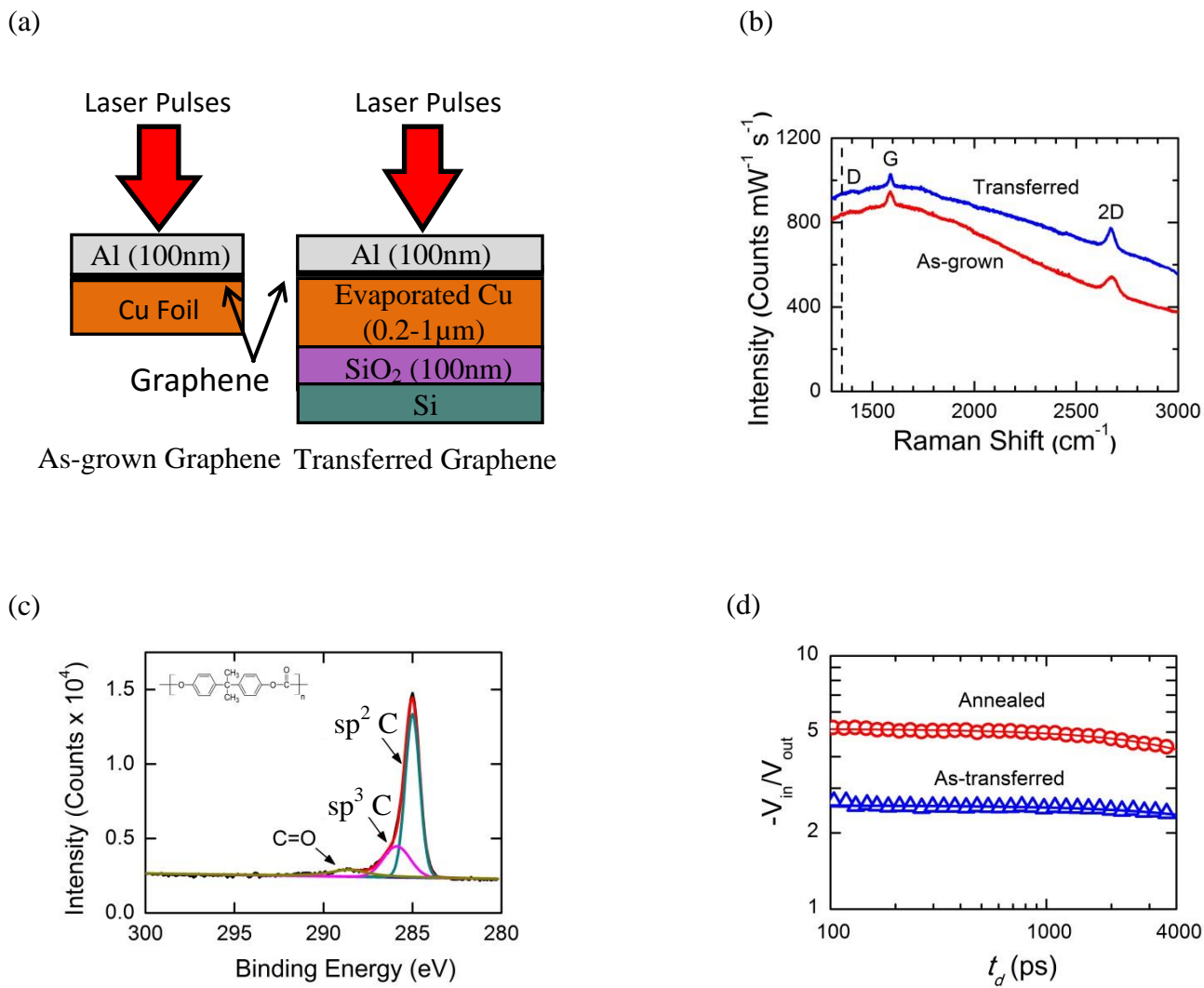

**Figure 1: (a)** A schematic diagram of the cross sectional view of our samples: (left) as-grown CVD graphene (grG) on copper foil (Set A), and (right) as-transferred graphene (trG) on thermal evaporated copper films (Set B and Set C). **(b)** Raman spectra of as-grown graphene on Cu foil and transferred graphene on evaporated Cu, as labeled, acquired using a 532nm continuous wave laser. **(c)** C1s core level spectra of transferred graphene on evaporated Cu. The binding energy of the $sp^2$ carbon bond is assigned at 285eV. $sp^3$ carbon and carbonyls C=O sub-peaks are fitted using a mixture of Gaussian and Lorentzian functions. The ratio of the integrated intensity of the C=O peak (residue) to that of the $sp^2$ carbon peak (graphene) is 0.07, comparable to that observed in as-grown graphene on Cu foil, indicating a low amount

of PC residues. Insert is the chemical formula for poly(bisphenol A carbonate) that is used as the supporting layer for the transfer. **(d)** Ratios of in-phase and out-of-phase TDTR signals as a function of the delay time for an as-transferred (open triangles) and an annealed (open circles) graphene samples, as labeled. The solid lines are calculations of a thermal model. We derive the thermal conductance of the graphene interfaces by fitting the calculations to the measurements.

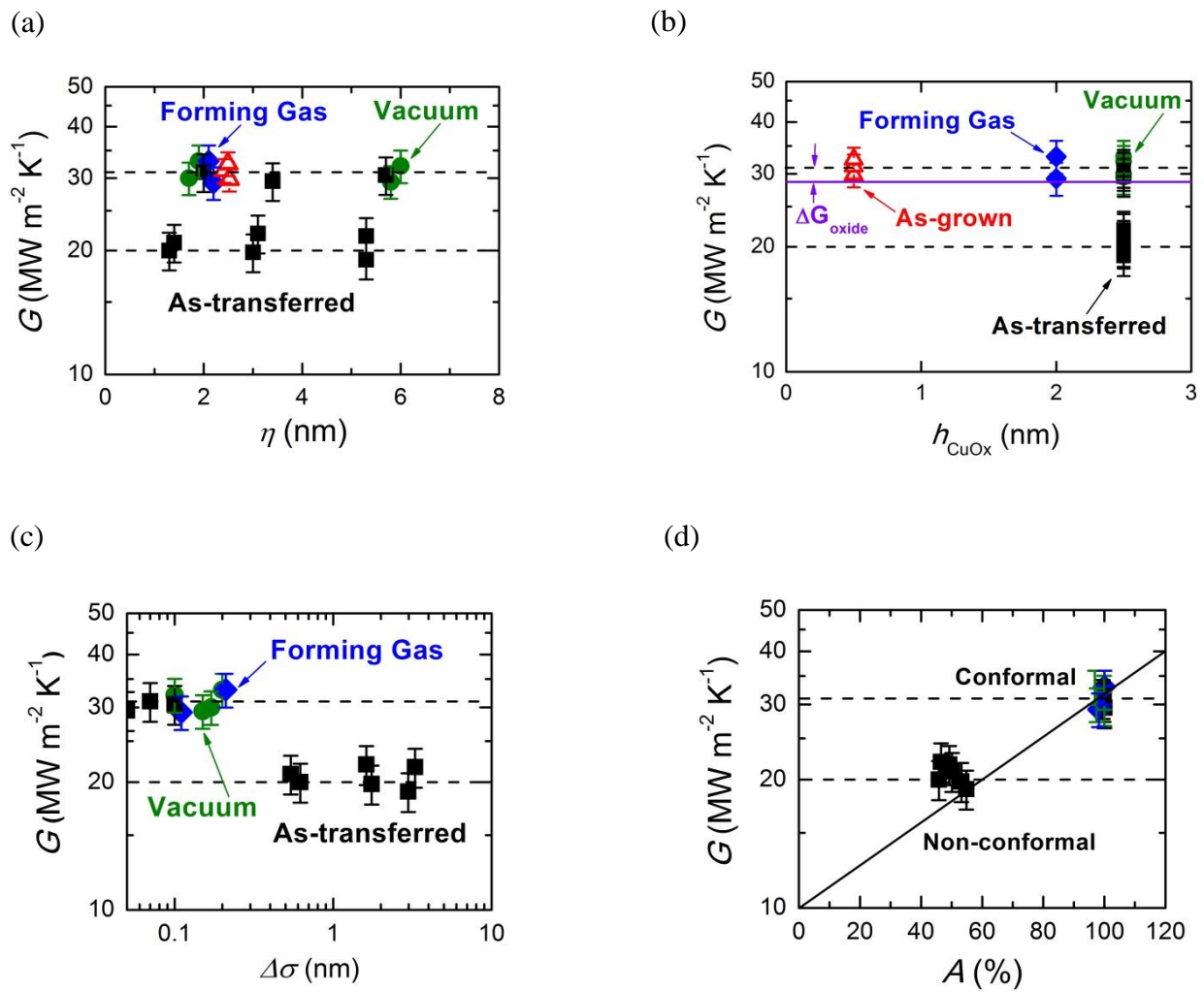

**Figure 2:** Thermal conductance $G$ of Al/Graphene/Cu interfaces for as-grown CVD graphene (Set A, open triangles), as-transferred graphene (Set B, solid squares), and transferred

graphene annealed at 300°C (Set C) either in vacuum (solid circles) or in forming gas (solid diamonds). The measurements are plotted as a function of **(a)** root mean square roughness ($\eta$) of the graphene derived from the AFM images; **(b)** thickness of the native oxide ($h_{CuOx}$) estimated using XPS; and **(c)** $\Delta\sigma$, which corresponds to degree of conformity of the transferred graphene, see the main text for the definition of $\Delta\sigma$. (d) area of contact ($A$) in percentage. The dashed lines at $G = 31$ MW m$^{-2}$ K$^{-1}$ and $G = 20$ MW m$^{-2}$ K$^{-1}$ are two distinct average values of $G$ of the Al/trG/Cu interfaces of the as-transferred graphene. The solid line in (b) is the expected $G$ when the thermal resistance of the 2.5 nm copper oxide is added to $G = 31$ MW m$^{-2}$ K$^{-1}$. The solid line in (d) represents the value of $G$ of Al/trG/Cu interfaces if $G$ is proportional to the area of contact.

(a)

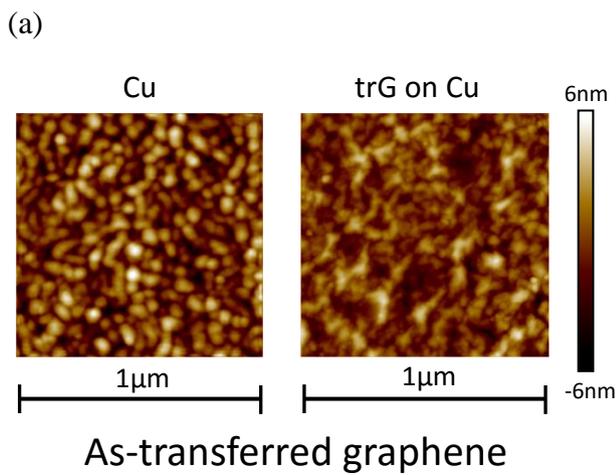

As-transferred graphene

(b)

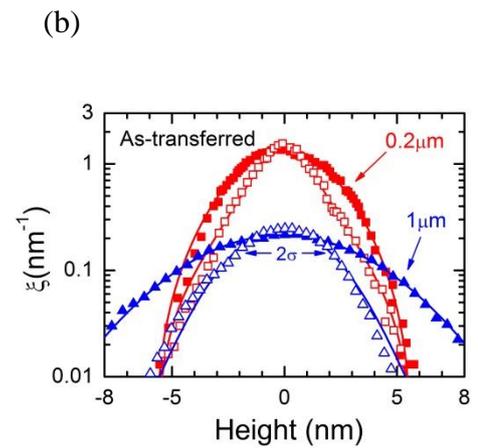

(c)

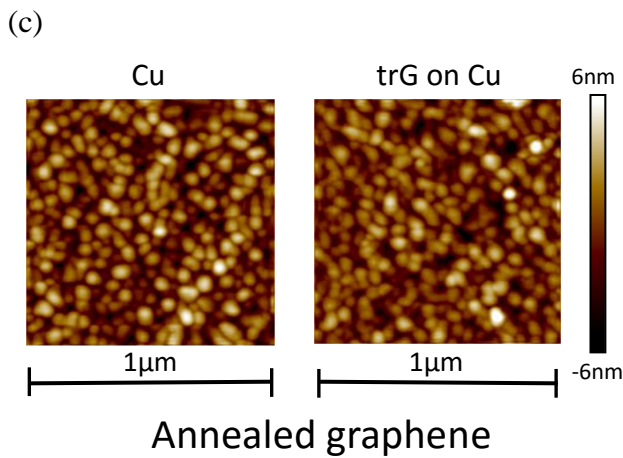

Annealed graphene

(d)

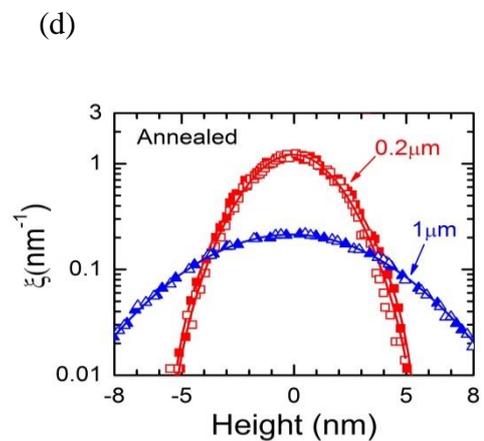

**Figure 3: (a), (c)** AFM images of evaporated Cu on SiO$_2$/Si substrate (left) and transferred graphene on evaporated Cu (right) of **(a)** an as-transferred graphene sample from Set B and **(c)** a transferred sample annealed in forming gas from Set C. The thickness of the evaporated Cu film is 0.2 μm for both samples. **(b), (d)** Depth histograms of evaporated Cu (solid symbols) and transferred graphene on evaporated Cu (open symbols) of (b) as-transferred and (d) annealed graphene samples. The thickness of the evaporated Cu film is 0.2 μm (squares, the same samples as in (a) and (c)) and 1 μm (triangles), respectively, as labeled. The spatial frequency (ξ), see the text for the definition, is plotted against the height. The depth histograms are fitted with a Gaussian function to obtain the standard deviation (σ) of the height distribution.

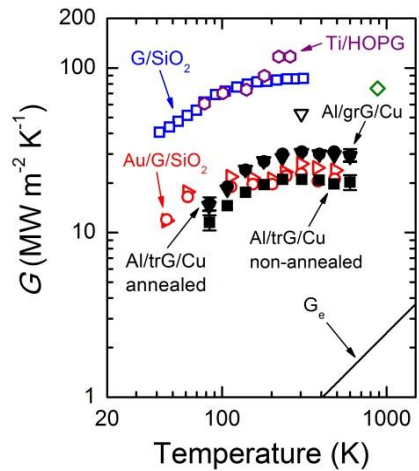

**Figure 4:** Temperature dependence of the thermal conductance $G$ of Al/grG/Cu interfaces of an as-grown CVD graphene (solid down triangles, this work), Al/trG/Cu interface of an as-transferred graphene (solid squares, this work), and Al/trG/Cu interface of an annealed graphene (solid circles, this work), compared to the thermal conductance of interfaces of CNT/SiO$_2$ (open diamond, ref 44), Au/Ti/graphite (open down triangle, ref 7), Ti/HOPG

(open hexagonal, ref 45), G/SiO2 (open square, ref 15), and Au/G/SiO$_2$ (open right triangles and open circles, ref 7). The solid line is the electronic thermal conductance of the Al/graphene/Cu interfaces estimated from the electrical resistance of the interface.